\documentclass[pra,twocolumn,showpacs,showkeys]{revtex4}
\usepackage{amsmath,amsgen,amstext,amsbsy,amsopn,amsthm,amssymb,epsf}
\usepackage{amssymb,graphics,graphicx,wrapfig}

\usepackage{amssymb,epsf}

\def\be{\begin{equation}}
\def\ee{\end{equation}}
\def\ba{\begin{align}}
\def\ea{\end{align}}
\def\bsplit{\begin{split}}
\def\esplit{\end{split}}
\def\bm{\begin{multline}}
\def\eem{\end{mutline}}
\def\bfig{\begin{figure}[htb]}
\def\efig{\end{figure}}
\renewcommand{\leq}{\;\leqslant\;}
\renewcommand{\geq}{\;\geqslant\;}
\newcommand{\dd}{{\rm d}}
\newcommand{\e}[1]{\,{\rm e}^{#1}\,}

\newcommand{\sumtwo}[2]{\sum_{\substack{#1 \\ #2}}}

\def\Tr{{\operatorname{Tr\,}}}

\newcommand{\upchi}{\raise 2pt \hbox{$\chi$}}
\newcommand{\caS}{{\mathcal S}}

\newcommand{\bbR}{{\mathbb R}}\newcommand{\bbZ}{{\mathbb Z}}

\newcommand{\bsx}{{\boldsymbol x}}\newcommand{\bsy}{{\boldsymbol y}}\newcommand{\bsz}{{\boldsymbol z}}

\begin{document}

\title[Critical Temperature of a Bose gas]{Critical Temperature of Dilute Bose Gases}

\author{Volker Betz and Daniel Ueltschi}
\affiliation{Department of Mathematics, University of Warwick, Coventry, CV4 7AL, United Kingdom}
\email{v.m.betz@warwick.ac.uk, daniel@ueltschi.org}

\begin{abstract}
We compute the critical temperature of Bose-Einstein condensation in dilute three-dimensional homogeneous Bose gases. Our method involves the models of spatial permutations and it should be exact to lowest order in the scattering length of the particle interactions. We find that the change in the critical temperature, $\Delta T_{\rm c} / T_{\rm c}^{(0)}$, behaves as $c a \rho^{1/3}$ with $c = -2.33$; this contradicts the current consensus among physicists.
\end{abstract}

\keywords{Dilute Bose gas, Bose-Einstein condensation} \pacs{05.70.Fh, 03.75.Hh, 05.30.Jp}

\maketitle

{\sc 1. Introduction.}
The study of the effects of particle interactions on the critical temperature of the Bose-Einstein condensation has a long and tortuous history. Several studies found that repulsive interactions decrease the critical temperature \cite{Fey,FW,Toy}, but most studies pointed to an increase of the critical temperature \cite{GKW,Hua,Sto,HGL,HK,AM,KPS,Kas,NL}. The current consensus among physicists is that the change $\Delta T_{\rm c} = T_{\rm c}^{(a)} - T_{\rm c}^{(0)}$ behaves as
\be
\frac{\Delta T_{\rm c}}{T_{\rm c}^{(0)}} \approx c a \rho^{1/3},
\ee
with positive constant $c \approx 1.3$ \cite{AM,KPS,Kas,NL}. Here, $T_{\rm c}^{(a)}$ denotes the critical temperature of a dilute homogeneous Bose gas and $a$ is the scattering length of the (repulsive) potential that describes the particle interactions. These results are reviewed in Refs \cite{BBHLV,Bla,SU}, where more details and additional references can be found. The latter article \cite{SU} proposes a mathematically rigorous upper bound on the interacting critical density; it is useful but it does not settle the question.

The present article uses a different approach that involves ``spatial random permutations". We initially expected that our calculations would confirm the consensus among physicists, but they rather seem to invalidate it: We find a negative constant $c = -\frac{8\pi}{3 \zeta(3/2)^{4/3}} \approx -2.33$, which implies that interactions decrease the critical temperature. While not rigorous, we expect our approach to be exact and to yield the correct constant. But we also trust the physics literature, and we are left puzzled. We do not know how to resolve these contradictions.

\medskip
{\sc 2. Overview of the method.}
We start with the Feynman-Kac representation of the Bose gas, where quantum particles become winding Brownian bridges in one more dimension \cite{Fey,Gin}. For dilute gases, Bose-Einstein condensation is signaled by the occurrence of infinite loops \cite{Suto,Uel1}. We actually consider the simpler models of spatial random permutations where Brownian bridges have been integrated \cite{BU1}. We derive a simplified model where the original two-jump interactions are replaced by ``cycle weights". We argue that the new model has almost the same marginal distribution on permutations, and the same critical temperature to lowest order in the scattering length. The critical temperature of the model with cycle weights can be computed explicitly. Many of the concepts and methods in this article can be fully justified mathematically. There are five instances, however, where a leap of faith is required. We have outlined these steps by writing them as ``conjectures". While the validity of the conjectures 1, 2, 4, and 5, seems clear, the conjecture 3 is tricky. If our calculations do not give the correct result, the most likely reason is the failure of this conjecture.

Finally, we put our method to a test by computing the free energy of the effective model. As it turns out, it is equal to the free energy of the dilute Bose gas \cite{Sei,Yin}, to leading order in $a$. This does not prove the method right, but this raises our confidence in it.

\medskip
{\sc 3. Mathematical setting.}
The state space for $N$ bosons in a box $\Lambda \subset \bbR^3$ is the Hilbert space $L^2_{\rm sym}(\Lambda^N)$ of square-integrable complex symmetric functions. The Hamiltonian is the Schr\"odinger operator
\be
H = -\sum_{i=1}^N \Delta_i + \sum_{1\leq i<j\leq N} U(x_i-x_j),
\ee
with $\Delta_i$ the Laplacian for the $i$-th particle, and $U$ a function that acts as a multiplication operator. We always suppose that $U$ is nonnegative, spherically symmetric, with finite range. It is characterized by its scattering length $a$. If $U$ is small and integrable, $a \approx \frac1{8\pi} \int U(x) \dd x$ is the first Born approximation to the scattering length. If $U$ consists of a hard core, $a$ is the radius of the hard core. We refer to Appendix C of the monograph \cite{LSSY} for the general definition and more information.

\medskip
{\sc 4. Feynman-Kac representation.}
The Feynman-Kac formula allows to represent quantum particles by Brownian bridges \cite{Fey,Gin}. The partition function of the Bose gas can be written as that of a model of ``spatial permutations" \cite{BU1}, namely,
\be
\begin{split}
Z &= \Tr_{L^2_{\rm sym}(\Lambda^N)} \e{-\beta H} \\
&= \frac1{N! (4\pi\beta)^{3N/2}} \sum_{\pi \in \caS_N} \int_{\Lambda^N} \dd\bsx \e{-H_1(\bsx,\pi)},
\end{split}
\ee
where $\caS_N$ denotes the set of permutations of $N$ elements, $\bsx = (x_1,\dots,x_N) \in \Lambda^N$, and where the ``Gibbs factor" $\e{-H_1(\bsx,\pi)}$ is given by
\bm
\label{eq for H1}
\e{-H_1(\bsx,\pi)} = \Bigl[ \prod_{i=1}^N (4\pi\beta)^{3/2} \int\dd W^{2\beta}_{x_i,x_{\pi(i)}}(\omega_i) \Bigr] \\
\exp \Bigl\{ -\frac12 \sum_{1\leq i<j\leq N} \int_0^{2\beta} U(\omega_i(s)-\omega_j(s)) \dd s \Bigr\}.
\end{multline}
In this article, $W^t_{x,y}$ denotes the Wiener measure for the Brownian bridge traveling from $x$ to $y$ in time $t$. It satisfies
\be
\int\dd W_{x,y}^t(\omega) = g_t(x-y),
\ee
with $g_t$ the Gaussian function, namely,
\be \label{plain gauss}
g_t(x) = (2\pi t)^{-3/2} \e{-x^2/2t}.
\ee
We temporarily ignore boundary effects, but we will need to take them into account later when we calculate the weights of macroscopic cycles.

Let $H_0(\bsx,\pi)$ denote the Hamiltonian of the model of spatial permutations that corresponds to the ideal Bose gas:
\be
H_0(\bsx,\pi) = \frac1{4\beta} \sum_{i=1}^N |x_i - x_{\pi(i)}|^2.
\ee
The equation \eqref{eq for H1} for $H_1$ can then be written as
\bm
\e{-H_1(\bsx,\pi)} = \e{-H_0(\bsx,\pi)} \Bigl[ \prod_{i=1}^N \int\dd\widehat W^{2\beta}_{x_i,x_{\pi(i)}}(\omega_i) \Bigr] \\
\exp \Bigl\{ -\frac12 \sum_{1\leq i<j\leq N} \int_0^{2\beta} U(\omega_i(s)-\omega_j(s)) \dd s \Bigr\}.
\end{multline}
Here, we introduced the normalized Wiener measure $\widehat W^t_{x,y}$, that is equal to $g_t^{-1}(x-y) W^t_{x,y}$.

\medskip
{\sc 5. Cycle lengths and Bose-Einstein condensation.}
The order parameter for Bose-Einstein condensation is Penrose and Onsager's ``off-diagonal long-range order" \cite{PO}. But following S\"ut\H o \cite{Suto}, we rather consider the length of permutation cycles. It is expected that Bose-Einstein condensation is accompanied by the occurrence of infinite cycles.

\smallskip
{\it CONJECTURE 1. The critical temperature for Bose-Einstein condensation is identical to the critical temperature for the occurrence of infinite cycles.}
\smallskip

This conjecture is compatible with the conclusions drawn in Ref.\ \cite{Uel1}; it was indeed argued that it is valid for dilute systems, but that it may be invalid otherwise.
From now on, we rover in the realm of spatial permutations, and we invoke methods and heuristics of classical statistical mechanics and of probability theory. The first step consists in deriving a model with ``two-jump interactions".

\medskip
{\sc 6. Model with 2-jump interactions.}
We consider the Hamiltonian
\be
H_2(\bsx,\pi) = H_0(\bsx,\pi) + \sum_{1\leq i<j\leq N} V_{ij}(\bsx,\pi)
\ee
where $V_{ij}(\bsx,\pi)$ depends on $x_i, x_{\pi(i)}, x_j, x_{\pi(j)}$, and it represents the interactions between the jumps $x_i \mapsto x_{\pi(i)}$ and $x_j \mapsto x_{\pi(j)}$. It involves the particle interactions and it is given by the formula
\be
\label{2-jump interactions}
V_{ij}(\bsx,\pi) = \int [1 - \e{-\frac14 \int_0^{4\beta} U(\omega(s)) \dd s}] \dd\widehat W^{4\beta}_{x_i-x_j, x_{\pi(i)} - x_{\pi(j)}}(\omega).
\ee
The general properties of $V_{ij}$ are not immediately apparent. By the Feynman-Kac formula, it is related to the integral kernel of the operator $\e{2\beta\Delta} - \e{\beta(2\Delta-U)}$. Let us notice that, when $U$ is a hard core potential of radius $a$, $V_{ij}$ is equal to the probability that a Brownian bridge from $x_i-x_j$ to $x_{\pi(i)} - x_{\pi(j)}$ intersects the ball of radius $a$ around the origin.

The formula \eqref{2-jump interactions} was computed in Ref.\ \cite{Uel2}, where it was also shown that, in a certain sense,
\be
\label{H1 vs H2}
H_1(\bsx,\pi) = H_2(\bsx,\pi) + O(a^2).
\ee
An expression for $O(a^2)$ can be found in \cite{Uel2} that results from a cluster expansion. Rigorous estimates for this term have yet to be derived, though.

\smallskip
{\it CONJECTURE 2. The critical temperature of the model $H_2$ is identical to that of the model $H_1$, up to a correction $o(a)$.}
\smallskip

While a bit simpler, the new model is still intractable {\it per se} and further simplifications are necessary. The next goal is a model of spatial permutations with ``cycle weights", see Eq.\ \eqref{Ham weights} below.

The correction $O(a^2)$ in Eq.\ \eqref{H1 vs H2} also involves the inverse temperature $\beta$ and the density $\rho$. One should discuss the dimensions that are present here. $H_1$ and $H_2$ are dimensionless; $a$ is like a length; $\beta$ is like the square of a length; $\rho^{-1}$ is like the cube of a length. It seems that the correction in \eqref{H1 vs H2} is really like $O((\beta\rho a)^2)$, which is dimensionless. It is not clear at this stage whether it is enough to suppose that $\beta \rho a \ll1$, or whether an additional condition is needed. In this article we always suppose that $\beta \rho^{2/3} \sim 1$, i.e.\ the system is in the regime of the phase transition. We only characterize the error terms $O(\cdot)$ and $o(\cdot)$ by their dependence on the scattering length $a$.

\medskip
{\sc 7. Using spatial averaging.}
We now describe the key step that allows to replace the 2-jump interactions $V_{ij}(\bsx,\pi)$ by cycle weights $\alpha_\ell$ within cycles of length $\ell$, and $\alpha_{\ell,\ell'}$ between cycles of length $\ell$ and $\ell'$. Cycle weights are much simpler because they do not depend on spatial positions. The difficulty is to make this simplification while keeping the critical density unchanged, at least to lowest order.

Let $\theta$ be a random variable that depends only on the permutation, not on the positions. Its expectation can be written as
\be
\begin{split}
E(\theta) &= \frac1{Z_2} \int\dd\bsx \sum_\pi \theta(\pi) \e{-H_2(\bsx,\pi)} \\
&= \frac1{Z_2} \sum_\pi \theta(\pi) Z^{(\pi)} \int\dd\mu^{(\pi)}(\bsx) \e{-\sum_{i<j} V_{ij}(\bsx,\pi)}.
\end{split}
\ee
We introduced the probability measure $\mu^{(\pi)}$ that depends on $\pi$, and which is defined by
\be
\label{def mu pi}
\dd\mu^{(\pi)}(\bsx) = \frac1{Z^{(\pi)}} \e{-H_0(\bsx,\pi)} \dd\bsx,
\ee
with
\be
Z^{(\pi)} = \int \e{-H_0(\bsx,\pi)} \dd\bsx.
\ee

We now introduce a ``permutation free energy" by
\be
\e{-F(\pi)} = \int\dd\mu^{(\pi)}(\bsx) \e{-\sum_{i<j} V_{ij}(\bsx,\pi)}.
\ee
The main step of our method is to take the integral inside the exponential, namely
\be
\label{new permutation free energy}
{} \e{-F(\pi)} \approx \exp\Bigl\{ - \int\dd\mu^{(\pi)}(\bsx)\sum_{i<j} V_{ij}(\bsx,\pi) \Bigr\}.
\ee
The left side is larger than the right side by Jensen's inequality, but it is hard to formulate a converse inequality. Let us carefully justify Eq.\ \eqref{new permutation free energy}. 
We need to take the large size of the system into account, and to think in terms of typical positions. For a given permutation, let $\bsy$ be a typical realization of the measure $\mu^{(\pi)}$, and let $\bsz$ be a typical realization of the measure
\[
\frac1{\rm normalization} \e{-H_0(\bsx,\pi)} \e{-\sum_{i<j} V_{ij}(\bsx,\pi)}.
\]
For all macroscopic observables $A$ we expect that
\be
A(\bsy) = A(\bsz) \, (1 + O(a)).
\ee
This holds in particular when the macroscopic observable is $\sum_{i<j} V_{ij}(\bsx,\pi)$. We have $\e{-F(\pi)} \approx \e{-\sum_{i<j} V_{ij}(\bsz,\pi)}$. We replace it by $\e{-F(\pi)} \approx \e{-\sum_{i<j} V_{ij}(\bsy,\pi)}$, the difference of free energies should be of order $O(a^2)$. The latter gives the expression \eqref{new permutation free energy}.

We get the new Hamiltonian
\be
\label{def H3}
H_3(\bsx,\pi) = H_0(\bsx,\pi) + \sum_{i<j} \int V_{ij}(\bsy,\pi)  \, \dd\mu^{(\pi)}(\bsy).
\ee

\smallskip
{\it CONJECTURE 3. The critical temperature of the model $H_3$ is identical to that of the model $H_2$, up to a correction $o(a)$.}
\smallskip

\medskip
{\sc 8. Model with cycle weights.}
The expression \eqref{def H3} can be simplified further. If $i$ and $j$ belong to the same cycle $\gamma$ of length $\ell$, we have
\be
\int V_{ij}(\bsy,\pi)  \, \dd\mu^{(\pi)}(\bsy) = \int V_{ij}(\bsy,\gamma)  \, \dd\mu^{(\gamma)}(\bsy),
\ee
where the new expectation refers to a system of $\ell$ points with the cyclic permutation $\gamma$. If $i$ and $j$ belong to different cycles $\gamma$ and $\gamma'$ of respective length $\ell$ and $\ell'$, we have
\be
\int V_{ij}(\bsy,\pi) \, \dd\mu^{(\pi)}(\bsy) = \int V_{ij}(\bsy,\gamma\cup\gamma')  \, \dd\mu^{(\gamma\cup\gamma')}(\bsy),
\ee
where $\gamma\cup\gamma'$ denotes the permutation of $\ell+\ell'$ elements, with two cycles of length $\ell$ and $\ell'$. Let us introduce
\begin{align}
&\alpha_\ell = \sum_{i,j \in \gamma, i<j} \int V_{ij}(\bsy,\gamma)  \, \dd\mu^{(\gamma)}(\bsy), \label{def one-cycle weight}\\
&\alpha_{\ell,\ell'} = \tfrac12 \sum_{i \in \gamma, j \in \gamma'} \int V_{ij}(\bsy,\gamma\cup\gamma')  \, \dd\mu^{(\gamma\cup\gamma')}(\bsy) \label{def two-cycle weight}.
\end{align}
Notice that these expressions depend on the cycle lengths, but not on the explicit cycles. We get a Hamiltonian for spatial permutations with ``cycle weights", namely
\be
\label{Ham weights}
\begin{split}
H_3(\bsx,\pi) = &H_0(\bsx,\pi) + \sum_{\ell\geq1} \bigl( \alpha_\ell - \alpha_{\ell,\ell} \bigr) r_\ell(\pi) \\
&+ \sum_{\ell,\ell' \geq 1} \alpha_{\ell,\ell'} r_\ell(\pi) r_{\ell'}(\pi).
\end{split}
\ee
An important remark is that the weights $\alpha_\ell$ and $\alpha_{\ell,\ell'}$ depend on the domain $\Lambda$ of the system. For the small cycles, this dependence is anecdotal as it disappears in the infinite volume limit. For macroscopic cycles, confinement in a bounded domain is important; since the cycles cover the domain, all points find themselves at finite (spatial) distance of other points that are very far away along the cycle. We need not worry about cycles of intermediate size, since these are known to involve only a vanishing fraction of particles \cite{Suto, BU1, BU2}.

\medskip
{\sc 9. Simplified 2-jump interaction.}
To first order, the 2-jump interaction is given by
\be
\begin{split}
V_{ij}(\bsx,\pi) &\doteq \frac14 \int_0^{4\beta} \dd s \int\dd\widehat W^{4\beta}_{x_i-x_j, x_{\pi(i)} - x_{\pi(j)}}(\omega) U(\omega(s)) \\
&\equiv V(x_i-x_j, x_{\pi(i)} - x_{\pi(j)}).
\end{split}
\ee
The notation $\doteq$ means that the equality holds to first order in $a$.
Recall that the first Born approximation to the scattering length is $\frac1{8\pi} \int U(x) \dd x \doteq a$. By the property of the Wiener measure,
\be
\int\dd\widehat W^{4\beta}_{x,y}(\omega) U(\omega(s)) = \int\dd z U(z) \frac{g_s(x-z) g_{4\beta-s}(y-z)}{g_{4\beta}(x-y)}.
\ee
It follows that
\be
V(x,y) \doteq 2\pi a \int_0^{4\beta} \frac{g_s(x) g_{4\beta-s}(y)}{g_{4\beta}(x-y)} \dd s.
\ee
The following formula about Gaussian functions is not hard to establish, and it is useful here:
\be
\frac{g_s(x) g_{t-s}(y)}{g_t(x-y)} = t^3 g_{st(t-s)} \bigl( (t-s)x + sy \bigr).
\ee
Then, after a change of variables,
\be
\label{simple V}
V(x,y) \doteq 8\pi\beta a \int_0^1 g_{4\beta s (1-s)} \bigl( (1-s)x + sy \bigr) \dd s.
\ee

\medskip
{\sc 10. Weights of finite cycles.}
We now calculate the one-cycle weights to first order in $a$. We do it first for finite cycles, which will lead to the result in Eq.\ \eqref{formula alpha}, and we will do it in the next paragraph for macroscopic cycles.
Recall the definition \eqref{def one-cycle weight} of the weight $\alpha_\ell$. We have
\be
\alpha_\ell = \tfrac12 \ell \sum_{j=2}^\ell \int V_{1j}(\bsy,\gamma) \dd\mu^{(\gamma)}(\bsy).
\ee
Gaussian functions satisfy $g_s * g_t = g_{s+t}$. We use this equation repeatedly, in conjunction with \eqref{def mu pi} and \eqref{simple V}, and we get
\be
\alpha_\ell \doteq (4\pi\beta\ell)^{5/2} a \sum_{j=2}^\ell \int_0^1 F_{\ell j}(s) \dd s,
\ee
with
\bm
\label{def F}
F_{\ell j}(s) = \frac1{|\Lambda|} \int_{\Lambda^4} \dd x_1 \dd x_2 \dd x_j \dd x_{j+1} \, g_{2\beta}(x_1 - x_2) \\
g_{2\beta (j-2)}(x_2 - x_j) g_{2\beta}(x_j - x_{j+1}) g_{2\beta (\ell-j)}(x_{j+1} - x_1) \\
g_{4\beta s(1-s)} \bigl( (1-s) (x_1-x_j) + s (x_2-x_{j+1}) \bigr).
\end{multline}
We used the fact that $Z^{(\gamma)} = |\Lambda| g_{2\beta\ell}(0) = |\Lambda| (4\pi\beta\ell)^{-3/2}$. We have neglected boundary effects, but they are irrelevant for finite $\ell$. We now consider the case $j=2$ and $\ell\geq3$. We actually have
\bm
F_{\ell 2}(s) = \frac1{|\Lambda|} \int_{\Lambda^3} \dd x_1 \dd x_2 \dd x_3 \, g_{2\beta}(x_1 - x_2) \\
g_{2\beta}(x_2 - x_3) g_{2\beta (\ell-2)}(x_3 - x_1) \\
g_{4\beta s(1-s)} \bigl( (1-s) (x_1-x_2) + s (x_2-x_1) \bigr).
\end{multline}
Using translation invariance, we can set $x_2=0$, picking up a factor $|\Lambda|$. We get
\bm
F_{\ell 2}(s) = (4\pi\beta)^{-6} (\ell-2)^{-3/2} \bigl( 2s(1-s) \bigr)^{-3/2} \int \dd x_1 \dd x_3 \\
\exp\Bigl\{ -\frac{x_1^2}{4\beta} - \frac{x_3^2}{4\beta} - \frac{(x_3-x_1)^2}{4\beta(\ell-2)} -\frac{((1-s)x_1 - s x_3)^2}{8\beta s(1-s)} \Bigr\}.
\end{multline}
The space variables factorize with respect to space directions. Integrating the Gaussians, we get
\be
F_{\ell 2}(s) = (4\pi\beta)^{-6} (\ell-2)^{-3/2} \bigl( 2s(1-s) \bigr)^{-3/2} \Bigl( \frac{4\pi\beta}{\sqrt{\det A}} \Bigr)^3
\ee
with the matrix $A$ given by
\be
A = \left( \begin{matrix} 1 + \frac1{\ell-2} + \frac{1-s}{2s} & -\frac12 - \frac1{\ell-2} \\ -\frac12 - \frac1{\ell-2} & 1 + \frac1{\ell-2} + \frac s{2(1-s)} \end{matrix} \right).
\ee
The determinant of this matrix is equal to $(2s (1-s))^{-1} (1 + \frac1{\ell-2})$. We obtain $F_{\ell 2}(s) = (4\pi\beta)^{-3} (\ell-1)^{-3/2}$. Let us note that it does not depend on $s$. This result holds for $\ell\geq3$ and for $j=2$, and also for $j=\ell$ which is the same. We now treat the case $\ell\geq3$ and $2<j<\ell$. We introduce $z_1 = x_1 - x_j$ and $z_2 = x_2 - x_{j+1}$ in \eqref{def F}; we replace $x_2$ by 0 using translation invariance, and we set $x_1 = x$. We obtain
\bm
F_{\ell j}(s) = (4\pi\beta)^{-15/2} \bigl( (j-2) (\ell-j) \bigr)^{-3/2} \bigl( 2s(1-s) \bigr)^{-3/2} \\
\int \dd x \dd z_1 \dd z_2 \exp\Bigl\{ -\frac{x^2}{4\beta} - \frac{(z_1-x)^2}{4\beta (j-2)} - \frac{(x-z_1+z_2)^2}{4\beta} \\
- \frac{(x+z_2)^2}{4\beta (\ell-j)} -\frac{((1-s)z_1 - s z_2)^2}{8\beta s(1-s)} \Bigr\} \\
= (4\pi\beta)^{-15/2} \bigl[ (j-2) (\ell-j) (2s(1-s)) \bigr]^{-3/2} \Bigl( \frac{(4\pi\beta)^{3/2}}{\sqrt{\det A}} \Bigr)^3.
\end{multline}
Here, the matrix $A$ is given by (with $u = 1 + \frac1{j-2}$ and $v = 1 + \frac1{\ell-j}$)
\be
A = \left( \begin{matrix} u+v & -u & v \\ -u & v + \frac{1-s}{2s} & -\frac12 \\ v & -\frac12 & v + \frac s{2(1-s)} \end{matrix} \right).
\ee
One easily finds $\det A = uv / 2s (1-s)$. It follows that
\be
F_{\ell j}(s) = (4\pi\beta)^{-3} (j-1)^{-3/2} (\ell-j+1)^{-3/2}.
\ee
The formula for $\ell\geq3$ and $j=2,\ell$ turns out to be compatible with the formula above. We have checked that it also holds in the case $\ell=j=2$. Notice that $F_{\ell j}(s)$ does not depend on $s$. After minor rearrangements, we obtain
\be
\label{formula alpha}
\alpha_\ell \doteq \frac{\ell a}{(4\pi\beta)^{1/2}} \sum_{j=1}^{\ell-1} \Bigl( \frac\ell{j(\ell-j)} \Bigr)^{3/2}.
\ee

It is not hard to see that
\be
\label{limit alpha}
\lim_{\ell\to\infty} \frac{\alpha_\ell}\ell \doteq \frac{2 \zeta(\frac32) a}{(4\pi\beta)^{1/2}} = 8 \pi \beta \rho_{\rm c}^{(0)} a.
\ee
This represents the energy per particle in a large finite cycle.

\medskip
{\sc 11. Weights of macroscopic cycles.}
When the cycles are macroscopic, i.e.\ $\ell \sim N$, we can no longer assume all points to be 
far away from the boundary of $\Lambda$; boundary effects have to be taken into account. For the Bose gas with 
periodic boundary conditions, \eqref{plain gauss} has to be replaced with the periodized version 
\be
g_t^{(\Lambda)}(x) = \sum_{y \in \bbZ^3} g_{t}(x + Ly).
\ee
Here, $L$ denotes the size of the box.
The identity  $g_t^{(\Lambda)} * g_s^{(\Lambda)} = g_{t+s}^{(\Lambda)}$ still holds, where the convolution is now 
defined by $f * g(x) = \int_{\Lambda} f(x-y) g(y) \, \dd y$. Thus \eqref{def F} holds with 
$g$ replaced by $g_{\Lambda}$ throughout, and indeed this is the correct finite volume expression even for finite
cycles. 
For $\ell$ finite and $\Lambda \to \infty$, we have 
$g_{2 \beta \ell}^{(\Lambda)}(x) \to g_{2 \beta \ell}(x)$, which justifies our using $g$ instead of $g^{(\Lambda)}$ in 
$\eqref{def F}$. But for $\ell = \varepsilon N$ with any $\varepsilon >0$, it is not hard to see that 
$|\Lambda| g_{2 \beta \ell}^{(\Lambda)}(x) \to 1$ for all $x$. Those Gaussians with macroscopic variance in 
\eqref{def F} must be replaced by $|\Lambda|^{-1}$. A similar calculation as above then leads to the result
\be
\alpha_\ell \doteq 8 \pi \beta \rho_{\rm c}^{(0)} a \ell + \frac{4\pi\beta a \ell^2}{|\Lambda|} - C a \beta^{-1/2}.
\ee
The constant is equal to $C = (3 - \frac32 \gamma_{1/2}) / \sqrt\pi$, see Eq.\ \eqref{limit alpha prime}, but this term does not play any r\^ole.

\medskip
{\sc 12. Two-cycle weights.}
Recall the definition \eqref{def two-cycle weight} of the weight $\alpha_{\ell,\ell'}$. We have also made computations in this case, and we have found that
\be
\label{formula alpha2}
\alpha_{\ell,\ell'} \doteq \frac{4\pi\beta \ell \ell' a}{|\Lambda|}.
\ee
In retrospect, we could have guessed this formula: From the definition \eqref{def two-cycle weight}, we clearly have
\be
\alpha_{\ell,\ell'} = \tfrac12 \ell \ell' \int V_{1,\ell+1}(\bsy,\gamma\cup\gamma') \, \dd\mu^{(\gamma\cup\gamma')}(\bsy).
\ee
Retracing our steps back to the Feynman-Kac representation of the quantum model, we see that $\int V_{1,\ell+1} \dd\mu^{(\gamma\cup\gamma')}$ gives the average interaction between the first Brownian bridge of the first loop and the first Brownian bridge of the second loop. Let us fix the first loop, and let us fix the imaginary time at which the interactions are computed. The integral over the second loop gives $8\pi a / |\Lambda|$, after dividing by the normalization. This quantity does not depend on the first loop nor on the imaginary time. We get a factor $\beta$ with the time integration, but the integral over the first loop is canceled by the normalization. We then get \eqref{formula alpha2}.

Because $\alpha_{\ell,\ell'}$ is proportional to $\ell \ell'$, there is an important simplification:
\be
\sum_{\ell,\ell'\geq1} \alpha_{\ell,\ell'} r_\ell(\pi) r_{\ell'}(\pi) = \frac{4\pi\beta a N^2}{|\Lambda|},
\ee
so that this term is a constant in the canonical ensemble. It does not affect the probability distribution of permutations, hence the critical density. We keep it, however, in order to perform the test of the free energy.

Let us subtract the limit \eqref{limit alpha} in the one-cycle weights, and define
\be
\label{alpha'}
\alpha_\ell' = \frac{2\ell a}{(4\pi\beta)^{1/2}} \Bigl[ \tfrac12 \sum_{j=1}^{\ell-1} \Bigl( \frac\ell{j(\ell-j)} \Bigr)^{3/2} - \zeta(\tfrac32) \Bigr].
\ee
We now consider the Hamiltonian
\bm
H_4(\bsx,\pi) = \frac{4\pi\beta a N^2}{|\Lambda|} + 8\pi\beta \rho_{\rm c}^{(0)} a N \\
+ H_0(\bsx,\pi) + \sum_{\ell\geq1} \alpha_\ell' r_\ell(\pi).
\end{multline}
$H_4$ is essentially identical to $H_3$, we only replaced certain expressions by their lowest order approximations. We should emphasize, however, that the approximating of $\alpha_{\ell,\ell'}$ has substituted the interactions between different cycles by a constant, which is a huge simplification.

\smallskip
{\it CONJECTURE 4. The critical temperature of the model $H_4$ is identical to that of the model $H_3$, up to a correction $o(a)$.}
\smallskip

\medskip
{\sc 13. Properties of one-cycle weights.}
We now show that $\alpha_\ell' = Ca (1 + O(\ell^{-1/5}))$, see Eq.\ \eqref{limit alpha prime} below. We will actually not use this formula directly, but it helps gaining a better understanding of the method.

Let us simplify the notation a bit and introduce $\phi_\ell$ such that $\alpha_\ell' = 2 (4\pi\beta)^{-1/2} a \phi_\ell$. We have
\be
\label{def phi}
\phi_\ell = \ell \sum_{j=1}^{\ell/2} \Bigl[ \Bigl( \frac\ell{j (\ell-j)} \Bigr)^{3/2} - \frac1{j^{3/2}} \Bigr] - \ell \sum_{j>\ell/2} \frac1{j^{3/2}}.
\ee
We assumed here that $\ell$ is odd. For even $\ell$, we need to subtract a term $2^{3/2} / \sqrt\ell$, which is smaller than the error term. The last term in \eqref{def phi} is equal to $2^{3/2} \sqrt\ell + O(\ell^{-1/2})$. For the first term, let us first observe that
\bm
\lim_{\ell\to\infty} \sqrt\ell \sum_{j=1}^{\ell/2} \Bigl[ \Bigl( \frac\ell{j (\ell-j)} \Bigr)^{3/2} - \frac1{j^{3/2}} \Bigr] \\
= \int_0^{1/2} \Bigl[ \frac1{(1-x)^{3/2}} - 1 \Bigr] \frac{\dd x}{x^{3/2}}.
\end{multline}
Let $\varepsilon>0$. We have
\bm
\ell \sum_{j=\varepsilon\ell}^{\ell/2} \Bigl[ \Bigl( \frac\ell{j (\ell-j)} \Bigr)^{3/2} - \frac1{j^{3/2}} \Bigr] \\
= \sqrt\ell \int_\varepsilon^{1/2} \Bigl[ \frac1{(1-x)^{3/2}} - 1 \Bigr] \frac{\dd x}{x^{3/2}} + O((\varepsilon \ell)^{-1/2}).
\end{multline}
By making the change of variables $x = \sin^2 \theta$, and then using trigonometric identities, we find
\be
\int_\varepsilon^{1/2} \frac{\dd x}{x^{3/2} (1-x)^{3/2}} = \frac{2-4\varepsilon}{\sqrt{\varepsilon (1-\varepsilon)}}.
\ee
One also has $\int_\varepsilon^{1/2} x^{-3/2} \dd x = 2/\sqrt\varepsilon - 2^{3/2}$. It follows that
\be
\int_\varepsilon^{1/2} \Bigl[ \frac1{(1-x)^{3/2}} - 1 \Bigr] \frac{\dd x}{x^{3/2}} = 2^{3/2} - 3\sqrt\varepsilon + O(\varepsilon^{3/2}).
\ee
Next, it is not hard to check that
\bm
\ell \sum_{j=1}^{\varepsilon\ell} \Bigl[ \Bigl( \frac\ell{j (\ell-j)} \Bigr)^{3/2} - \frac1{j^{3/2}} \Bigr] = \tfrac32 \sum_{j=1}^{\varepsilon\ell} \frac1{\sqrt j} + O(\varepsilon^2 \ell) \\
= 3\sqrt{\varepsilon\ell} - 3 + \tfrac32 \gamma_{1/2} + O(\varepsilon^2 \ell) + O((\varepsilon\ell)^{-1/2}),
\end{multline}
where $\gamma_{1/2}$ is a generalized Euler constant, that is defined by
\be
\gamma_{1/2} = \lim_{n\to\infty} \Bigl[ \sum_{j=1}^n \frac1{\sqrt j} - \int_1^n \frac{\dd x}{\sqrt x} \Bigr] = 0.5396...
\ee
Collecting all the necessary terms, and choosing $\varepsilon = \ell^{-3/5}$, we obtain
\be
\phi_\ell = -3 + \tfrac32 \gamma_{1/2} + O(\ell^{-1/5}).
\ee
It follows that
\be
\label{limit alpha prime}
\alpha_\ell' = - (6 - 3\gamma_{1/2}) (4\pi\beta)^{-1/2} a (1 + O(\ell^{-1/5})).
\ee
We observe that the weights $\alpha_\ell'$ are negative. This is important, as it will lead to the decrease of the critical temperature. Notice that the sign of $\alpha_\ell'$ was not obvious prior to the computation.

\medskip
{\sc 14. The critical density with cycle weights.}
Let $H_5$ be like $H_4$ but without the constant terms. Explicitly,
\be
H_5 = H_0(\bsx,\pi) + \sum_{\ell\geq1} \alpha_\ell' r_\ell(\pi).
\ee
These Hamiltonians generate the same Gibbs distributions, so that they have the same critical density. We now calculate the critical density of the model $H_5$ using the grand-canonical ensemble. Namely, we consider the pressure
\be
p(\beta,\mu) = \frac1{\beta|\Lambda|} \log \!\! \sum_{N\geq0} \frac{\e{\beta\mu N}}{(4\pi\beta)^{\frac{3N}2} N!} 
\sum_{\pi \in \caS_N} \int_{\Lambda^N} \!\! \dd\bsx \e{-H_5(\bsx,\pi)}.
\ee
It can be computed by going to the Fourier space and by introducing occupation numbers, as for the usual ideal Bose gas. In the limit of infinite volumes, we find
\be
\label{pressure H5}
p(\beta,\mu) = \sum_{\ell\geq1} \frac{\e{\beta\mu\ell - \alpha_\ell'}}{(4\pi)^{3/2} (\beta \ell)^{5/2}}.
\ee
The pressure is finite for $\mu\leq0$ and it is increasing in $\mu$. It is infinite for $\mu>0$. As in the ideal gas, the critical density is given by the derivative of $p$ at $\mu=0$:
\be
\label{crit dens}
\rho_{\rm c} = \frac\partial{\partial\mu} p(\beta,\mu) \Big|_{\mu=0-} = \sum_{\ell\geq1} \frac{\e{-\alpha_\ell'}}{(4\pi\beta \ell)^{3/2}}.
\ee

We need to emphasize that this formula for the critical density is computed by analogy with the ideal Bose gas and its relevance for macroscopic cycles is far from clear. In the case $\alpha_j \equiv 0$ (where $\rho_{\rm c} = \rho_{\rm c}^{(0)}$), S\"ut\H o proved that it is indeed the critical density for the occurrence of infinite cycles \cite{Suto}. This result was extended to the case of vanishing cycle weights --- such that $\alpha_\ell' \to 0$ faster than $1 / \log\ell$ as $\ell\to\infty$ in Ref.\ \cite{BU2}. The weights found in this article are small (they are proportional to $a$) but they converge to a constant, as we saw in the paragraph above. We use the formula here nonetheless.

\smallskip
{\it CONJECTURE 5. The critical density of the model $H_5$ is given by Eq.\ \eqref{crit dens}.}
\smallskip

\medskip
{\sc 15. The critical density.}
The critical density is given by Eq.\ \eqref{crit dens}, where the cycle weights are given by Eq.\ \eqref{alpha'}. To first order,
\be
\label{diff crit dens}
\rho_{\rm c}^{(a)} - \rho_{\rm c}^{(0)} \doteq -\frac{2a}{(4\pi\beta)^2} \sum_{\ell\geq1} \frac1{\ell^{1/2}} \Bigl[ \tfrac12 \sum_{j=1}^{\ell-1} \Bigl( \frac\ell{j(\ell-j)} \Bigr)^{3/2} - \zeta(\tfrac32) \Bigr].
\ee
It turns out that the sums can be explicitly calculated. We have
\be
\rho_{\rm c}^{(a)} - \rho_{\rm c}^{(0)} \doteq -\frac{2a}{(4\pi\beta)^2} \lim_{L\to\infty} C_L
\ee
with
\be
C_L = \sum_{\ell=1}^L \frac1{\ell^{1/2}} \Bigl[ \tfrac12 \sum_{j=1}^{\ell-1} \Bigl( \frac\ell{j(\ell-j)} \Bigr)^{3/2} - \zeta(\tfrac32) \Bigr].
\ee
Interchanging the sums over $\ell$ and $j$, we get
\be
\label{CL}
\begin{split}
C_L &= \sum_{j\geq1} \frac1{j^{3/2}} \Bigl[ \tfrac12 \sum_{\ell=j+1}^L \frac\ell{(\ell-j)^{3/2}} - \sum_{\ell=1}^L \frac1{\ell^{1/2}} \Bigr] \\
&= \sum_{j \geq 1} \frac1{j^{3/2}} \Bigl[ \tfrac12 \sum_{\ell=1}^{L-j} \frac{\ell+j}{\ell^{3/2}} - \sum_{\ell=1}^{L-j} \frac1{\ell^{1/2}} \Bigr] \\
&\hspace{2cm} - \sum_{j \geq 1} \frac1{j^{3/2}} \sum_{\ell = (L-j)_+ + 1}^L \frac1{\ell^{1/2}}.
\end{split}
\ee
The first line of the latter expression is equal to
\be
\begin{split}
&\tfrac12 \sum_{j\geq1} \frac1{j^{3/2}} \sum_{\ell=1}^{L-j} \Bigl( \frac j{\ell^{3/2}} - \frac1{\ell^{1/2}} \Bigr) \\
&=\tfrac12 \sumtwo{j,\ell\geq1}{j+\ell \leq L} \Bigl( \frac1{j^{1/2} \ell^{3/2}} - \frac1{j^{3/2} \ell^{1/2}} \Bigr) = 0.
\end{split}
\ee
We are left with the last line of Eq.\ \eqref{CL}. Splitting the sum over $j$, we get
\be
\label{CL 12}
\begin{split}
C_L &= -\sum_{j=1}^L \frac1{j^{3/2}} \sum_{\ell=L-j+1}^L \frac1{\ell^{1/2}} - \sum_{j \geq L+1} \frac1{j^{3/2}} \sum_{\ell=1}^L \frac1{\ell^{1/2}} \\
&\equiv - C_L^{(1)} - C_L^{(2)}.
\end{split}
\ee
After a change of variables, the first term is given by
\be
C_L^{(1)} = \frac1{L^2} \sum_{j=1}^L \frac1{(\frac jL)^{3/2}} \sum_{\ell=1}^j \frac1{(1 - \frac jL + \frac\ell L)^{1/2}}.
\ee
As $L\to\infty$ we get a Riemann integral, namely,
\bm
\lim_{L \to \infty} C_L^{(1)} = \int_0^1 \frac{\dd x}{x^{3/2}} \int_0^x \frac{\dd y}{(1-x+y)^{1/2}} \\
= 2 \int_0^1 \frac{1 - \sqrt{1-x}}{x^{3/2}} \dd x = -4 + 4 \int_0^1 \frac{\dd x}{\sqrt{x(1-x)}}.
\end{multline}
The last identity follows from an integration by parts. The substitution $x = \sin^2 \theta$ eventually gives
\be
\lim_{L \to \infty} C_L^{(1)} = 2\pi - 4.
\ee
The second term in \eqref{CL 12} is equal to
\be
C_L^{(2)} = \bigg( \frac1L \sum_{j\geq1} \frac1{(1 + \frac j\ell)^{3/2}} \biggr) \biggl( \frac1L \sum_{\ell=1}^L \frac1{(\frac\ell L)^{1/2}} \biggr).
\ee
Then
\be
\lim_{L \to \infty} C_L^{(2)} = \int_0^\infty \frac{\dd x}{(1+x)^{3/2}} \int_0^1 \frac{\dd y}{\sqrt y} = 4.
\ee
We have obtained that $\lim_L C_L = -2\pi$, and so we have from \eqref{diff crit dens} that
\be
\label{voila b}
\frac{\rho_{\rm c}^{(a)} - \rho_{\rm c}^{(0)}}{\rho_{\rm c}^{(0)}} \doteq \frac{2\sqrt\pi}{\zeta(\frac32)} a \beta^{-1/2}.
\ee

\medskip
{\sc 16. The critical temperature.}
We have found the constant $b$ for the linear change in the critical density, i.e.
\be
\frac{\rho_{\rm c}^{(a)} - \rho_{\rm c}^{(0)}}{\rho_{\rm c}^{(0)}} \doteq b a \beta^{-1/2}.
\ee
We would rather know the constant $c$ for the linear change in the critical temperature, i.e.
\be
\frac{T_{\rm c}^{(a)} - T_{\rm c}^{(0)}}{T_{\rm c}^{(0)}} \doteq c a \rho^{1/3}.
\ee
The critical line can be seen as a manifold in the space $(\rho,\beta,a)$; at $a=0$ we have the relation
\be
\label{relation at a=0}
\rho = \frac{\zeta({\frac32})}{(4\pi\beta)^{3/2}}.
\ee
Recall that
\be
\label{3D manifold}
\frac{\partial a}{\partial\rho} \frac{\partial\rho}{\partial\beta} \frac{\partial\beta}{\partial a} = -1.
\ee
We have
\begin{align}
&\frac{\partial a}{\partial\rho} \doteq \Bigl( \frac{\rho_{\rm c}^{(a)} - \rho_{\rm c}^{(0)}}a \Bigr)^{-1} \doteq \frac1{b \beta^{-\frac12} \rho}, \\
&\frac{\partial\rho}{\partial\beta} = -\frac32 \frac{\zeta(\frac32)}{(4\pi)^{3/2} \beta^{5/2}}, \\
&\frac{\partial\beta}{\partial a} = -\frac1{T^2} \frac{\partial T}{\partial a} \doteq -\frac1{T^2} c \rho^{1/3} T = -c \beta \rho^{1/3}.
\end{align}
Inserting these expressions into \eqref{3D manifold}, and using \eqref{relation at a=0}, we find
\be
\label{relation bc}
c = -\frac{4 \, \pi^{1/2}}{3 \, \zeta(\frac32)^{1/3}} b.
\ee

We can now compute the constant for the change in the critical temperature.
Using $b = 2\sqrt\pi / \zeta(\frac32)$ from Eq.\ \eqref{voila b} and the relation above, we finally obtain
\be
c = -\frac{8\pi}{3 \zeta(3/2)^{4/3}} \approx -2.33.
\ee
The negative sign means that interactions discourage Bose-Einstein condensation. Our finding is in contradiction with the physicists' consensus $c \approx 1.3$.

\medskip
{\sc 17. A test of the method.}
We now compute the free energy of our approximate model, and we compare it with the formula for the dilute Bose gas, which is known explicitly \cite{Sei,Yin}.

The free energy of the model $H_4$ is defined by
\be
f(\beta,\rho) = \frac1{\beta|\Lambda|} \log \frac1{(4\pi\beta)^{\frac{3N}2} N!} \sum_{\pi \in \caS_N} \int_{\Lambda^N} \dd\bsx \e{-H_4(\bsx,\pi)}.
\ee
It is related to the free energy $\hat f$ of the model $H_5$ by
\be
\label{energie libre}
f(\beta,\rho) = 4\pi a \rho^2 + 8 \pi a \rho_{\rm c}^{(0)} \rho + \hat f(\beta,\rho).
\ee
The existence of the thermodynamic limit and the equivalence of ensembles were shown in Ref.\ \cite{BU2}. Then $\hat f$ is given by the Legendre transform of the corresponding pressure,
\be
\hat f(\beta,\rho) = \sup_\mu \bigl[ \rho\mu - p(\beta,\mu) \bigr] \equiv \rho\mu^* - p(\beta,\mu^*).
\ee
The pressure for $H_5$ has already been computed, the result can be found in Eq.\ \eqref{pressure H5}. To first order,
\be
p(\beta,\mu) \doteq p^{(0)}(\beta,\mu) - \sum_{\ell\geq1} \frac{\e{\beta\mu\ell} \alpha_\ell'}{(4\pi)^{3/2} (\beta \ell)^{5/2}}.
\ee
The equation for $\mu^*$ is $\rho = \partial p / \partial\mu$; for $\rho \leq \rho_{\rm c}$,
\be
\rho \doteq \rho^{(0)}(\beta,\mu^*) - \sum_{\ell\geq1} \frac{\e{\beta\mu^* \ell} \alpha_\ell'}{(4\pi\beta \ell)^{3/2}}.
\ee
Notice that $\mu^*=0$ when $\rho \geq \rho_{\rm c}$.
We have
\be
\label{on avance}
\hat f(\beta,\rho) = \rho\mu^* - p^{(0)}(\beta,\mu^*) + \sum_{\ell\geq1} \frac{\e{\beta\mu^* \ell} \alpha_\ell'}{(4\pi)^{3/2} (\beta\ell)^{5/2}}.
\ee
Since $\mu^*$ differs from the maximizer of $\rho\mu - p^{(0)}$ by $O(a)$, we have
\be
\rho\mu^* - p^{(0)}(\beta,\mu^*) = f^{(0)}(\beta,\rho) + O(a^2).
\ee
We now calculate the last term of \eqref{on avance}. Using the expression \eqref{alpha'} of $\alpha_\ell'$, we get
\bm
\sum_{\ell\geq1} \frac{\e{\beta\mu^* \ell} \alpha_\ell'}{(4\pi)^{3/2} (\beta\ell)^{5/2}} = \frac{2a}{\beta (4\pi\beta)^2} \\
\times \biggl[ \tfrac12 \sum_{\ell\geq1} \sum_{j=1}^{\ell-1} \frac{\e{\beta\mu^* j} \e{\beta\mu^* (\ell-j)}}{j^{3/2} (\ell-j)^{3/2}} - \zeta(\tfrac32) \sum_{\ell\geq1} \frac{\e{\beta\mu^* \ell}}{\ell^{3/2}} \biggr].
\end{multline}
The first term of the right side is equal to $4\pi a \rho^{(0)}(\beta,\mu^*)^2$, and the second term is equal to $8 \pi a \rho_{\rm c}^{(0)} \rho^{(0)}(\beta,\mu^*)$. We have
\be
\rho^{(0)}(\beta,\mu^*) = \begin{cases} \rho + O(a) & \text{if } \rho \leq \rho_{\rm c}, \\ \rho_{\rm c} + O(a) & \text{if } \rho \geq \rho_{\rm c}. \end{cases}
\ee
Then
\be
\hat f(\beta,\rho) \doteq f^{(0)}(\beta,\rho) + \begin{cases} 4\pi a \rho (\rho - 2 \rho_{\rm c}) & \text{if } \rho \leq \rho_{\rm c} \\ -4\pi a \rho_{\rm c}^2 & \text{if } \rho \geq \rho_{\rm c} \end{cases}
\ee
Together with \eqref{energie libre}, we get
\be
f(\beta,\rho) \doteq f^{(0)}(\beta,\rho) + 4\pi a [2\rho^2 - (\rho-\rho_{\rm c})_+^2].
\ee
The right side is indeed equal to the free energy of the dilute Bose gas up to a correction $O(a^2)$. This formula was rigorously established recently \cite{Sei,Yin}.

The test of the free energy does not prove the validity of our method, but it seems fairly robust nonetheless. It would detect a mistake in the definition or in the calculation of the weights. We can also be confident that no important term was left behind when the model was progressively simplified. It would be interesting to apply it to the other methods that served to compute the constant $c$. Our understanding would be greatly enhanced if we knew which methods pass it and which fail to pass it.

\medskip
{\sc 18. Conclusion.}
The main goal of this article is to propose a method that allows to compute the effects of interactions on the critical temperature of the dilute Bose gas. We used the concept of ``spatial random permutations", whose main advantage is to allow the heuristics and the intuition of statistical mechanics. Our findings contradict those of our colleagues from physics and this casts a shadow on the method. We have been careful to state our conjectures precisely, which will help resolve the discrepancy in the future. We have also performed a test (the calculation of the free energy) and the method passed it.
Besides the main goal, our approach suggests several models that may be of interest as effective models, even if they do not describe the dilute Bose gas exactly.

\medskip {\footnotesize {\bf Acknowledgments:} We thank Markus Holzmann and Robert Seiringer for useful discussions and comments. D.U. is grateful for the hospitality of the University of Arizona and of CNRS Marseille, where parts of this project were carried forward. V.B.\ is supported by the EPSRC fellowship EP/D07181X/1 and D.U. is supported in part by the grant DMS-0601075 of the US National Science Foundation.  }

\end{document}